\documentclass[a4paper,11pt]{article}
\usepackage{jheppub}
\usepackage{graphicx,amsmath,amssymb,dcolumn,bm}
\usepackage{fixmath}
\usepackage{feynmp}
\usepackage{epic,eepic}
\usepackage{slashed}
\allowdisplaybreaks[3]
\usepackage{calrsfs}
\DeclareMathAlphabet{\pazocal}{OMS}{zplm}{m}{n}

\title{Twist-2 relation and sum rule for tensor-polarized 
parton distribution functions of spin-1 hadrons}
\author[a,b]{S. Kumano}
\author[c]{and Qin-Tao Song}
\affiliation[a]{KEK Theory Center,
             Institute of Particle and Nuclear Studies, \\
             High Energy Accelerator Research Organization (KEK),\\
             Oho 1-1, Tsukuba, Ibaraki, 305-0801, Japan}
\affiliation[b]{J-PARC Branch, KEK Theory Center,
             Institute of Particle and Nuclear Studies, KEK, \\
           and Theory Group, Particle and Nuclear Physics Division, 
           J-PARC Center, \\
           Shirakata 203-1, Tokai, Ibaraki, 319-1106, Japan}
\affiliation[c]{School of Physics and Microelectronics, Zhengzhou University, \\
             Zhengzhou, Henan 450001, China}
\emailAdd{shunzo.kumano@kek.jp}
\emailAdd{songqintao@zzu.edu.cn}
\abstract{
Sum rules for structure functions and their twist-2 relations 
have important roles in constraining their magnitudes and 
$x$ dependencies and in studying higher-twist effects.
The Wandzura-Wilczek (WW) relation 
and the Burkhardt-Cottingham (BC) sum rule
are such examples for the polarized structure functions
$g_1$ and $g_2$. Recently, new twist-3 and twist-4 parton distribution
functions were proposed for spin-1 hadrons, so that it became possible
to investigate spin-1 structure functions 
including higher-twist ones.
We show in this work that an analogous twist-2 relation and a sum rule
exist for the tensor-polarized parton distribution functions $f_{1LL}$ 
and $f_{LT}$, where $f_{1LL}$ is a twist-2 function
and $f_{LT}$ is a twist-3 one.
Namely, the twist-2 part of $f_{LT}$ is expressed by an integral of
$f_{1LL}$ (or $b_1$) and the integral of the function
$f_{2LT} = (2/3) f_{LT} -f_{1LL}$ over $x$ vanishes. 
If the parton-model sum rule for $f_{1LL}$ ($b_1$) is applied 
by assuming vanishing tensor-polarized antiquark distributions, 
another sum rule also exists for $f_{LT}$ itself.
These relations should be valuable for studying tensor-polarized
distribution functions of spin-1 hadrons and for separating twist-2
components from higher-twist terms, as the WW relation and
BC sum rule have been used for investigating $x$ dependence and 
higher-twist effects in $g_2$.
In deriving these relations, we indicate that four twist-3 multiparton
distribution functions $F_{LT}$, $G_{LT}$, $H_{LL}^\perp$, and $H_{TT}$
exist for tensor-polarized spin-1 hadrons.
These multiparton distribution functions are also interesting
to probe multiparton correlations in spin-1 hadrons.
In the near future, we expect that physics of spin-1 hadrons will become 
a popular topic, since there are experimental projects 
to investigate spin structure of the spin-1 deuteron 
at the Jefferson Laboratory, the Fermilab, 
the nuclotron-based ion collider facility,
the electron-ion colliders in US and China
in 2020's and 2030's.
}

\begin{document} 
\maketitle\flushbottom

\section{Introduction}
\label{introduction}

High-energy spin physics has been one of exciting fields 
in hadron physics since the late 1980's
for clarifying the origin of nucleon spin.
In addition to longitudinally-polarized collinear structure functions,
we investigate transverse-spin and three-dimensional (3D) 
structure functions nowadays.
Furthermore, structure functions of spin-1 hadrons will be investigated
extensively in the near future due to the existence of 
new tensor-polarized observables.

We have been investigating structure functions of spin-1 hadrons,
especially on the spin-1 deuteron. 
There are four tensor-polarized structure functions
$b_{1-4}$ for a spin-1 hadron 
in charged-lepton deep inelastic scattering \cite{fs83,Hoodbhoy:1988am}.
The leading-twist functions $b_1$ and $b_2$ are related with each other
by the Callan-Gross type relation $2x b_1 = b_2$, and other functions
$b_3$ and $b_4$ are higher-twist ones. For finding the overall 
$x$-dependent functional form, there is a useful parton-model sum rule
for $b_1$ \cite{b1-sum}. The $b_1$ measurement by the HERMES collaboration
obtained a finite sum $\int dx b_1 \ne 0$ \cite{Airapetian:2005cb}, 
which indicated a finite tensor-polarized antiquark distribution.
Possible tensor-polarized parton distribution functions (PDFs)
were proposed for explaining the HERMES data 
\cite{tensor-pdfs}.
In future, an accurate determination of $b_1$ will be made 
by the experiment at the Thomas Jefferson National Accelerator 
Facility (JLab) \cite{jlab-b1}.
This $b_1$ project is interesting in the sense that 
the HERMES data are different from a conventional theoretical
estimate based on a convolution description \cite{b1-convolution}.
In addition, the gluon transversity $\Delta_T g$ exists 
for the spin-1 deuteron, although it does not exist 
for the spin-1/2 nucleons. 
This is also an interesting observable to find new hadron physics
in the deuteron beyond the simple bound system of nucleons.
It is expected to be measured by an JLab experiment
\cite{jlab-gluon-trans,ma-spin-1-2013}.
These $b_1$ and $\Delta_T g$ measurements could be continued
at the electron-ion collider (EIC) \cite{eic} and
the Electron-ion collider in China (EicC) \cite{eicc}.

Such tensor-polarized structure functions can be also investigated
at hadron accelerator facilities.
In fact, the proton-deuteron Drell-Yan process could probe
the tensor-polarized PDFs, especially the tensor-polarized
antiquark distributions \cite{pd-drell-yan,Kumano:2016ude}
and the gluon transversity \cite{ks-trans-g-2020}.
It could be realized as a Fermilab-SpinQuest experiment \cite{Fermilab-dy} 
when a polarized-deuteron target becomes ready \cite{Keller:2020wan}.
In addition, since the Nuclotron-based Ion Collider fAcility (NICA)
will have a polarized-deuteron beam, the tensor-polarized structure
functions and the gluon transversity will be investigated,
for example, by observing J/$\psi$ production \cite{nica}.
These structure functions are interesting for investigating especially
exotic aspects of hadron physics, such as hidden color 
\cite{miller-b1} and non-nucleonic component \cite{transversity-model}
in the deuteron.

On the other hand, transverse-momentum-dependent 
parton distribution functions (TMDs) 
became one of hot topics in hadron physics.
It is intended to understand not only basics 3D structure of hadrons
but also to find explicit color degrees of freedom in terms
of color flow. Because the color is confined in hadrons, it is not
easy to find its explicit signature in observables. 
The TMDs are such quantities to probe the color directly
\cite{ks-tmd-2021}.
For example, a color Aharonov-Bohm effect and color entanglement 
phenomena could be investigated by the TMDs.
In addition, gluon condensates are now investigated with
the understanding of gluon TMDs in the nucleons and nuclei.

Recently, we investigated the tensor-polarized TMDs for spin-1 hadrons 
and found 30 new TMDs at the twist 3 and twist 4 \cite{ks-tmd-2021}
in addition to the twist-2 functions \cite{bm-2000}.
Integrating the TMDs over the partonic transverse momentum,
we also found that there are three new collinear PDFs
at the twist 3 and twist 4. Therefore, including the twist-2 PDF,
we have the collinear PDFs $f_{1LL}$, $e_{LL}$, $f_{LT}$, and 
$f_{3LL}$ for spin-1 hadrons.
By considering this situation,
the purposes of this work are the following.
\begin{itemize}
\vspace{-0.25cm}
\setlength{\leftskip}{-0.28cm}
\setlength{\itemsep}{-0.11cm} 
\item[(1)] 
We derive a useful twist-2 relation and a sum rule
for $f_{1LL}$ and $f_{LT}$ in analogy to 
the Wandzura-Wilczek relation \cite{ww-1977}
and the Burkhardt-Cottingham sum rule \cite{bc-1970}
for the structure functions $g_1$ and $g_2$.
Here, $f_{1LL}$ is a twist-2 function and $f_{LT}$ is a twist-3 one,
and $f_{1LL}$ is often used as the tensor-polarized structure function $b_1$.
\item[(2)] 
We show that four twist-3 multiparton distribution functions
exist in a tensor-polarized spin-1 hadron.
\item[(3)] 
We show that the leading deviation of the twist-2 relation,
namely the higher-twist term, is expressed by these twist-3 multiparton 
distribution functions, so that
we try to obtain the full-decomposition expression
of $f_{LT}$ into twist-2 and twist-3 terms.
\end{itemize}
\vspace{-0.25cm}
\noindent
For simply deriving the twist-2 relation and sum rule, 
the studies of the twist-3 multiparton distributions functions
are not necessary. However, we investigate one more step
to express the higher-twist term of $f_{LT}$ by 
the twist-3 multiparton distributions functions
in this work.

In this paper, the Wandzura-Wilczek relation and Burkhardt-Cottingham 
sum rule are introduced by the operator-product-expansion formalism
in Sec.\,\ref{ww-bc-sum}.
Then, the details are explained on our derivations
of analogous relations for tensor-polarized PDFs
in Sec.\,\ref{sum-spin-1}.
First, correlation functions and collinear PDFs are introduced,
and the matrix element of a nonlocal vector operator is expressed
by the tensor-polarized collinear PDFs in 
Sec.\,\ref{collinear-pdfs}.
Next, the nonlocal operator is written in term of the gluon field tensor
and it is related to the multiparton distribution functions in 
Sec.\,\ref{twist-3-multiparton}.
A useful twist-2 relation and a sum rule are derived 
for $f_{1LL}$ and $f_{LT}$ in Sec.\,\ref{twist-2-relations}.
Our studies are summarized in Sec.\,\ref{summary}.

\section{\boldmath Wandzura-Wilczek relation and Burkhardt-Cottingham sum rule}
\label{ww-bc-sum}

In investigating structure functions, there are useful sum rules and 
relations among them for finding their functional behavior.
For example, there is a useful relation for the polarized structure 
function $g_2$, which exists in the spin-1/2 nucleons. 
Since we intend to derive a relation which is analogous to
the Wandzura-Wilczek (WW) relation and 
also a sum rule like the Burkhardt-Cottingham (BC) sum rule,
we introduce their outline within the formalism of 
operator product expansion.

The polarized distribution functions $g_{1L}$, $g_T$, and $g_{3L}$
are defined by the matrix element of a nonlocal operator as
\begin{align}
& 
\int \frac{d(P^+ \xi^-)}{2\pi} \, e^{ixP^+ \xi^-}
\langle \, P , S \left | \, \bar\psi  (0) \,  
\gamma^\mu \gamma_5 \psi  (\xi)  \, \right | P, \,  S \,
\rangle _{\xi^+ =0, \, \vec\xi_T=0} 
\nonumber \\
& \ \ \ \ \ 
= 2 M_N 
\left [ \, 
g_{1L} (x) \bar n^\mu S \cdot n + g_T (x) S_T^\mu
+ g_{3L} (x) \frac{M^2_N}{(P^+)^2} n^\mu S \cdot n \,
\right ] .
\label{eqn:g1-gT-g3}
\end{align} 
Here, $M_N$ is the nucleon mass,
$\psi$ is the quark field,
the lightcone vectors $n$ and $\bar n$ are defined by
\begin{align}
n^\mu =\frac{1}{\sqrt{2}} (\, 1,\, 0,\, 0,\,  -1 \, ), \ \ 
\bar n^\mu =\frac{1}{\sqrt{2}} (\, 1,\, 0,\, 0,\,  1 \, ) ,
\label{eqn:lightcone-n-nbar}
\end{align} 
$P$ and $S$ are the nucleon momentum and spin,
$S_T^\mu$ is the transverse-spin vector 
\cite{kt-1999,ks-tmd-2021}, $\xi$ is the space-time coordinate, 
the variable $x$ is the momentum fraction carried by a parton
and it is defined by $k^+ =x P^+$,
and the lightcone variables $a^\pm$ indicate
$a^\pm = (a^0 \pm a^3)/\sqrt{2}$.
In Eq.\,(\ref{eqn:g1-gT-g3}), the gauge link is abbreviated.
In this paper, the momentum and renormalization scale 
dependence ($Q^2$, $\mu^2$) is not explicitly written
in the PDFs and structure functions.

The moments of Eq.\,(\ref{eqn:g1-gT-g3}) become
\begin{align}
& 
\frac{1}{2 M_N (P^+)^{n-1}} \, n_{\mu_1} \cdots n_{\mu_{n-1}}
\langle \, P , S \left | \, 
R^{\sigma \{ \mu_1 \cdots \mu_{n-1} \}}
\, \right | P, \,  S \, \rangle
\nonumber \\
&
= \bar n^\sigma (S \cdot n) 
\int_{-1}^1 dx x^{n-1} g_{1L} (x)
+ S_T^\sigma \int_{-1}^1 dx x^{n-1} g_T (x) ,
\label{eqn:g1gt-moments-1}
\end{align} 
by keeping the terms up to twist 3. 
Hereafter, the twist-4 distribution function $g_{3L}$ 
and twist-4 terms are neglected in this section.
The structure functions are classified by the twist,
which is defined by the mass dimension minus spin,
in the operator product expansion
\cite{Anselmino-1995,Blumlein-1997-1999,Braun-2001,kt-1999,lr-2000,leader-book}.
The local operators $R^{\sigma \{ \mu_1 \cdots \mu_{n-1} \}}$
for describing the polarized structure functions are defined by
\begin{align}
R^{\sigma \{ \mu_1 \cdots \mu_{n-1} \}}
 = i^{n-1} \bar\psi \gamma^\sigma \gamma_5 
      D^{ \{ \mu_1} \cdots D^{\mu_{n-1} \} } \psi ,
\label{eqn:operator-R}
\end{align} 
where $D^\mu$ is the QCD covariant derivative given by
$D^\mu = \partial^\mu - ig A^\mu$
with the QCD coupling constant $g$ and the gluon field $A^\mu$, 
and the curly bracket $\{\ \}$ indicates the symmetrization
of all the Lorentz indices 
as defined in Eq.\,(\ref{eqn:twist-2-3}).
The higher-twist ($\ge 4$) trace terms 
$g_{\mu_i \mu_j} R^{\sigma \{ \mu_1 \cdots \mu_{n-1} \}}$
and $g_{\sigma \mu_i} R^{\sigma \{ \mu_1 \cdots \mu_{n-1} \}}$
should be subtracted to make the operator traeless;
however, they are not explicitly written 
in Eq.\,(\ref{eqn:operator-R}). 
The gluon field $A^\mu$ contains the SU(3) generator $t^a$, defined by
the Gell-Mann matrix $\lambda^a$ as $t^a = \lambda^a/2$
with the color index $a$, as $A^\mu = A^\mu_a t^a$.
Since these operators contain both twist-2 and twist-3 
components, they should be separated so as to have definite twists as
\cite{jaffe-ji-1991,jaffe-1996,Blumlein-1997-1999,kt-1999,Braun-2001}
\begin{align}
R^{\sigma \{ \mu_1 \cdots \mu_{n-1} \}}
& = R^{\{ \sigma \mu_1 \cdots \mu_{n-1} \}}
  + R^{[ \sigma \{ \mu_1 ] \cdots \mu_{n-1} \}} ,
\nonumber \\
R^{\{ \sigma \mu_1 \cdots \mu_{n-1} \}}
& = \frac{1}{n}
\left [  
  R^{ \sigma \{ \mu_1 \mu_2 \cdots \mu_{n-1} \}}
+ R^{ \mu_1  \{ \sigma \mu_2 \cdots \mu_{n-1} \}}
+ R^{ \mu_2  \{ \mu_1 \sigma \cdots \mu_{n-1} \}}
+ \cdots
\right ] ,
\nonumber \\
R^{[ \sigma \{ \mu_1 ] \cdots \mu_{n-1} \}}
& = \frac{1}{n}
\left [  
(n-1) R^{ \sigma \{ \mu_1 \mu_2 \cdots \mu_{n-1} \}}
- R^{ \mu_1  \{ \sigma \mu_2 \cdots \mu_{n-1} \}}
- R^{ \mu_2  \{ \mu_1 \sigma \cdots \mu_{n-1} \}}
- \cdots
\right ] ,
\label{eqn:twist-2-3}
\end{align} 
where the first term 
$R^{\{ \sigma \mu_1 \cdots \mu_{n-1} \}}$
has spin $n$ and the second one 
$R^{[ \sigma \{ \mu_1 ] \cdots \mu_{n-1} \}}$
has $n-1$ so that they are definite twist-2 and 
twist-3 operators.
The twist-3 operator $ R^{[ \sigma \{ \mu_1 ] \cdots \mu_{n-1} \}}$
is obtained by expanding the nonlocal operator
in the Taylor series as
\begin{align}
\xi_{\mu} \,
        \bar{\psi}(0) \left( \partial^\mu \gamma^{\sigma} 
                            - \partial^\sigma \gamma^\mu \right ) \gamma_5 \,
        \psi( \xi ) 
\Longrightarrow
\text{twist-3:} \ R^{[ \sigma \{ \mu_1 ] \cdots \mu_{n-1} \}} ,
\label{eqn:twist-3-derivative}
\end{align} 
where $\partial^\mu = \partial/\partial \xi_\mu$.
Then, it is written by the following
operators with the gluon field tensor $G^{\mu\nu}$,
the quark mass ($m_q$), and equation of motion as \cite{kt-1999}
\begin{align}
& \! \! \!
\xi_{\mu} \,
        \bar{\psi}(0) \left( \partial^\mu \gamma^\sigma
                            -\partial^\sigma \gamma^\mu \right )
        \gamma_5 \, 
         \psi( \xi )
= g \int^1_0 dt \, \bar{\psi}(0) 
   \left\{
          i \gamma_5 \left( t - \frac{1}{2} \right)  G^{\sigma\rho}(t\xi)
          - \frac{1}{2} 
          \tilde{G}^{\sigma\rho}(t\xi)  \right\} 
      \xi_{\rho} \slashed{\xi}  \psi (\xi)
\nonumber \\
& \ \hspace{3.0cm}
   + \,2  m_{q} \bar{\psi}(0) \gamma_5 \sigma^{\sigma\rho} \xi_{\rho} \psi (\xi)
\nonumber\\
& \ \hspace{3.0cm}
+ \,  \bar{\psi}(0)
            \gamma_5 \sigma^{\sigma\rho} \xi_{\rho} 
        ( i \slashed{D} - m_{q} ) \psi (\xi)  - \bar{\psi}(0)
          ( i \overleftarrow{\slashed{D}} + m_{q}) 
           \gamma_5 \sigma^{\sigma\rho} \xi_{\rho}  \psi (\xi),
\label{eqn:derivative-op}
\end{align}
where the dual field tensor $\tilde G^{\mu\nu}$ is defined by
$\tilde G^{\mu\nu} = \epsilon^{\mu\nu\rho\sigma} G_{\rho\sigma}/2$
\cite{Itzykson-Zuber-book}
with the convention $\epsilon^{0123}=+1$.
The first line of the right-hand side is from the quark-gluon-quark
correlation, the second one is the quark-mass term,
and the third one is the equation-of-motion term.
Scale evolution was studied by using these three types of operators; 
however, we do not step into such details.
Interested readers may read, for example, the summary article 
of Ref.\,\cite{kt-1999}.

In this way, the matrix elements of these operators are
generally expressed for the nucleons as 
\begin{align}
\langle \, P, S \, \big | \, 
R^{\{ \sigma \mu_1 \cdots \mu_{n-1} \}} \, 
\big | \, P, S \, \rangle 
& = \frac{2}{n} a_n M_N
\left [   
S^\sigma P^{\mu_1} \cdots P^{\mu_{n-1}}
+ S^{\mu_1} P^\sigma \cdots P^{\mu_{n-1}} \cdots
\right ],
\nonumber \\
\langle \, P, S \, \big | \, 
R^{[ \sigma \{ \mu_1 ] \cdots \mu_{n-1} \}} \, 
\big | \, P, S \, \rangle 
& = \frac{2}{n} d_n M_N
\big [
(S^\sigma P^{\mu_1} - S^{\mu_1} P^\sigma)
\, P^{\mu_2} \cdots P^{\mu_{n-1}} 
\nonumber \\
& \ \hspace{1.5cm} 
+ (S^\sigma P^{\mu_2} - S^{\mu_2} P^\sigma)
\, P^{\mu_1} \cdots P^{\mu_{n-1}} 
+ \cdots \big] ,
\label{eqn:matrix-twist-2-3}
\end{align} 
where $a_n$ and $d_n$ are constants to indicate the magnitudes
of the twist-2 and twist-3 matrix elements.
The polarized structure functions $g_1$ and $g_2$ are given by 
$g_{1L}$ and $g_T$ as
$g_1 (x) = [g_{1L} (x) + g_{1L} (-x)]/2$ and
$g_1 (x) + g_2 (x) = [g_{T} (x) + g_{T} (-x)]/2$.
From Eqs.\,(\ref{eqn:g1gt-moments-1}), (\ref{eqn:operator-R}),
(\ref{eqn:twist-2-3}), and (\ref{eqn:matrix-twist-2-3}), 
we obtain the moments as
\begin{align}
\int_0^1 dx x^{n-1} g_1 (x)
& = \frac{1}{2} a_n ,  
\nonumber \\
\int_0^1 dx x^{n-1} \left [ \, g_1 (x) +  g_2 (x) \, \right ]
& = \frac{1}{2}
\left ( \frac{1}{n} a_n + \frac{n-1}{n} d_n  \right ) .
\label{eqn:g1gt-moments}
\end{align} 
In these equations, we use $g_1$ and $g_2$ for a single flavor.
The second equation is written by using the first one as
\begin{align}
\int_0^1 dx x^{n-1} g_2 (x) 
= \int_0^1 dx x^{n-1} \left[ - g_1 (x) + \int_x^1 \frac{dy}{y} g_1 (y) \right ] 
 + \frac{n-1}{2n} d_n .
\label{eqn:g2-moments}
\end{align} 
From this equation, the structure function $g_2$ is
written in terms of twist-2 and twist-3 parts separately as
\begin{align}
g_2 (x) = g_2^{WW} (x) & + \bar g_2 (x), 
\label{eqn:g2-twist-2-3}
\\
g_2^{WW} (x) & = - g_1 (x) + \int_x^1 \frac{dy}{y} g_1 (y), 
\label{eqn:ww}
\\
\int_0^1 dx \, x^{n-1} \bar g_2 (x) & = \frac{n-1}{2n} d_n .
\label{eqn:g2-twist-3}
\end{align} 
Equation (\ref{eqn:ww}) is the WW relation which is valid
in the twist-2 level by neglecting higher-twist effects.
There is also a similar relation for the chiral-odd twist-3
structure function $h_L$ and the twist-2 one $h_1$
\cite{jj-1991,chiral-odd-twist-3}.
In addition, higher-twist terms are explicitly written
by multiparton distribution functions
\cite{chiral-odd-twist-3}.
If the WW relation of Eq.(\ref{eqn:ww}) is integrated over $x$,
it becomes
\begin{align} 
\int_0^1 dx \, g_2^{WW} (x) = 0 .
\label{eqn:bc-sum}
\end{align}
This relation is the BC sum rule, which was originally derived
by using the dispersion relation for the virtual Compton amplitude.
Here, the convergence of this sum could depend on 
the $x$-dependent functional form of $g_2$ at small $x$.
Furthermore, no operator is defined for $n=1$ 
in Eq.\,(\ref{eqn:operator-R}), so that 
the WW relation and BC sum rule could not be rigorously proven
in the operator-product-expansion formalism
\cite{Anselmino-1995,Blumlein-1997-1999,Braun-2001,br-book,
accardi-2009,Deur-2019}.
On the other hand, these relations are satisfied 
even if perturbative QCD corrections are included
in coefficient functions \cite{Braun-2001}.

\section{Parton distribution functions 
of spin-1 hadrons and their twist-2 relation and sum rule}
\label{sum-spin-1}

For studying structure functions, sum rules provide useful information
on their $x$-dependent functional forms.
There are sum rules for the structure functions and TMDs
of spin-1 hadrons \cite{b1-sum,ks-tmd-2021}.
If tensor-polarized antiquark distributions vanish,
there is a sum rule for the twist-2 collinear structure function
as $\int dx b_1 (x)=0$.
In addition, due to the time-reversal
invariance of collinear parton distributions, there exist
the sum rules
$ \int \! d^2 k_T h_{1LT}   (x, k_T^{\, 2})
= \! \int \! d^2 k_T g_{LT}  (x, k_T^{\, 2})
= \! \int \! d^2 k_T h_{LL}  (x, k_T^{\, 2})
= \! \int \! d^2 k_T h_{3LL} (x, k_T^{\, 2}) =0 $
for the TMDs of tensor-polarized spin-1 hadrons.
In this subsection, we show the existence of a new sum rule
and a twist-2 relation, which are analogous to 
the BC sum rule and the WW relation, respectively.

\subsection{Matrix elements of nonlocal operators
and parton distribution functions}
\label{collinear-pdfs}

The PDFs of hadrons are often discussed by correlation functions.
The PDFs of spin-1/2 nucleons are now theoretically 
investigated including higher-twist ones. On the other hand, 
the PDFs of spin-1 hadrons are not well studied 
especially for the tensor-polarized part. 
The TMDs and PDFs of spin-1 hadrons were investigated for 
the twist-2 in Ref.\,\cite{bm-2000}, and twist-3 and twist-4 functions
were recently proposed in Ref.\,\cite{ks-tmd-2021}.
The PDFs of hadrons are generally defined from
the correlation function 
\begin{align}
& \Phi_{ij}^{[c]} (k, P, T \, | \, n )  
= \int  \frac{d^4 \xi}{(2\pi)^4} \, e^{ i k \cdot \xi}
\langle \, P, T \left | \, 
\bar\psi _j (0) \,  W^{[c]} (0, \xi)  
 \psi _i (\xi)  \, \right | P, T \, \rangle ,
\label{eqn:correlation-q}
\end{align} 
which is related to the amplitude to extract a parton 
from a hadron and then to insert it into the hadron
at a different spacetime point $\xi$.
Here, $k$ is the quark momentum, 
the hadron momentum and tensor polarization are denoted 
by $P$ and $T$, 
respectively,
$W^{[c]}$ is the gauge link for satisfying the color gauge 
invariance, and $c$ indicates the integral path.
We do not write the spin vector polarization $S$ explicitly 
in Eq.\,(\ref{eqn:correlation-q}) 
because only the tensor polarization $T$,
which is specific to hadrons with spin$\ge 1$, is investigated
in this paper. The vector polarization part is essentially the same
as the one for the spin-1/2 nucleons.
From the general correlation function in Eq.\,(\ref{eqn:correlation-q}),
we obtain the collinear correlation function
by integrating it over the lightcone momentum $k^-$
and the transverse momentum $k_T$,
and fixing the $k^+$ component as
\begin{align}
\Phi_{ij} (x, P, T ) 
& = \int d^2 k_T \, dk^+ dk^- 
\Phi^{[c]}_{ij} (k, P, T \, |n) \, \delta (k^+  -x P^+) ,
\nonumber \\
& 
= \int  \frac{d\xi^-}{2\pi} \, e^{ixP^+ \xi^-}
\langle \, P , T \left | \, 
\bar\psi _j (0) \,  W (0, \xi \, |n)  
\psi _i (\xi)  \, \right | \! P, \,  T \,
\rangle _{\xi^+ =0, \, \vec\xi_T=0} .
\label{eqn:correlation-pdf}
\end{align}
Here, $W (a,b \, |n)$ indicates the gauge line connecting 
$a = (a^+, a^-, \vec a_T)$
to $b = (b^+, b^-, \vec b_T)$ 
along the straight lightcone direction of $\xi^-$.
Since the link is along the straight line, there is no 
path-$c$ dependence.


The TMDs and collinear PDFs
for spin-1 hadrons are defined in various traces 
of the TMD and collinear correlation function as $\text{Tr} (\Phi \Gamma )$, 
where $\Gamma$ is expressed by $\gamma$ matrices,
in Ref.\,\cite{ks-tmd-2021}.
In this subsection, the tensor-polarized collinear PDFs are studied and
the gauge link $W (0, \xi \, |n)$ is not explicitly written.
Later,  the Fock-Schwinger gauge $\xi_{\mu} A^{\mu}(\xi) =0$
\cite{bb-1988} is used from Sec.\,\ref{twist-3-multiparton}
and the gauge link is unity $W (0, \xi \, |n) =1$ 
in any case for the collinear PDFs.
However, this link appears in Sec.\,\ref{twist-3-multiparton},
for example in Eq.\,(\ref{eqn:gl3}),
for obtaining our WW- and BC-like relations 
because its total derivative does not vanish.
Here, we are interested in collinear PDFs which 
are defined in the trace as
\begin{align}
\Phi^{[\Gamma]} (x,P,T) & \equiv
\frac{1}{2} \, \text{Tr}
\left [ \, \Phi (x, P, T) \, \Gamma \, \right ]
= \frac{1}{2} \Phi_{ij} (x, P, T) \, (\Gamma)_{ji}
\nonumber \\
& = \frac{1}{2} \int  \frac{d\xi^-}{2\pi} \, e^{ixP^+ \xi^-}
\langle \, P , T \left | \, 
\bar\psi (0) 
\, \Gamma \, \psi (\xi)  \, \right | \! P, \,  T \,
\rangle _{\xi^+ =0, \, \vec\xi_T=0} .
\label{eqn:trace-correlation}
\end{align}
The tensor polarization $T^{\mu\nu}$ is generally expressed 
by the polarizations $S_{LL}$, $S_{LT}^\mu$, and $S_{TT}^{\mu\nu}$ as 
\cite{ks-trans-g-2020,ks-tmd-2021}
\begin{align}
T^{\mu\nu}  = \frac{1}{2} & \left [ \frac{4}{3} S_{LL} \frac{(P^+)^2}{M^2} 
               \bar n^\mu \bar n^\nu 
          - \frac{2}{3} S_{LL} ( \bar n^{\{ \mu} n^{\nu \}} -g_T^{\mu\nu} )
+ \frac{1}{3} S_{LL} \frac{M^2}{(P^+)^2}n^\mu n^\nu
\right.
\nonumber \\
& \ 
\left.
+ \frac{P^+}{M} \bar n^{\{ \mu} S_{LT}^{\nu \}}
- \frac{M}{2 P^+} n^{\{ \mu} S_{LT}^{\nu \}}
+ S_{TT}^{\mu\nu} \right ],
\label{eqn:spin-1-tensor-1}
\\[-0.90cm] \nonumber
\end{align}
where $a^{\{ \mu} b^{\nu \}}$ is the symmetrized combination
$a^{\{ \mu} b^{\nu \}} = a^\mu b^\nu + a^\nu b^\mu$.
The tensor-polarization parameters $S_{LL}$, $S_{LT}^\mu$, 
and $S_{TT}^{\mu\nu}$ are explained in Appendix of
Ref.\,\cite{bm-2000}. 
The $S_{LL}$ is associated with the tensor polarization
along the $z$ direction, $S_{TT}^{\mu\nu}$ is the linear polarization 
in the transverse plane, and $S_{LT}^\mu$ is the polarization
in the plane in-between.

We investigate possible relations of 
tensor-polarized PDFs $f_{1LL}$ and $f_{LT}$,
where the function $f_{1LL}$ is twist 2 and $f_{LT}$ is twist 3,
in analogy to the WW and BC relations for $g_1$ and $g_2$.
The function $f_{1LL}$ is defined in $\Phi^{[ \gamma^+ ]} (x, P, T)$, 
and $f_{LT}$ is in $\Phi^{[ \gamma^i ]} (x, P, T)$, 
and the actual expressions are given in the 
TMD form in Ref.\,\cite{ks-tmd-2021}.
The twist-4 function $f_{3LL}$ exists in another vector type correlation function
$\Phi^{[ \gamma^- ]} (x, P, T)$, so that it is also included 
in the following discussion of this subsection.
There is another twist-3 function $e_{LL}$ defined
in $\Phi^{[\mathbf{1}]} (x,P,T)$, where $\mathbf{1}$
is the $4\times 4$ identity matrix,
and it is also listed in this subsection.
There are related works on the TMDs and PDFs of spin-1 hadrons
\cite{Hoodbhoy:1988am,bm-2000,ma-spin-1-2013}.
Integrating the TMD expressions of Eqs.\,(33), (43), and (52)
in Ref.\,\cite{ks-tmd-2021} for 
$\Gamma=\gamma^+$, $\gamma^i$, $\gamma^-$, and $\mathbf{1}$
over the transverse momentum $\vec k_T$, we obtain
PDF expressions in terms of operator matrix elements as
\begin{align}
\! \!
\Phi^{[\gamma^+]} (x,P,T) & 
= \! \int \! \frac{d\xi^-}{4\pi} \, e^{ixP^+ \xi^-}
\langle \, P , T \left | \, 
\bar\psi (0) 
\, \gamma^+  \psi (\xi)  \, \right | \! P, \,  T \,
\rangle _{\xi^+ =0, \, \vec\xi_T=0}
= S_{LL} \, f_{1LL} (x), 
\nonumber \\
\! \!
\Phi^{[\gamma^\alpha]} (x,P,T) & 
= \! \int \! \frac{d\xi^-}{4\pi} \, e^{ixP^+ \xi^-}
\langle \, P , T \left | \, 
\bar\psi (0) 
\, \gamma^\alpha \, \psi (\xi)  \, \right | \! P, \,  T \,
\rangle _{\xi^+ =0, \, \vec\xi_T=0}
= \frac{M}{P^+}  S_{LT}^{\, \alpha} \, f_{LT} (x) ,
\nonumber \\
\! \!
\Phi^{[\gamma^-]} (x,P,T) & 
= \! \int \! \frac{d\xi^-}{4\pi} \, e^{ixP^+ \xi^-}
\langle \, P , T \left | \, 
\bar\psi (0) 
\, \gamma^-  \psi (\xi)  \, \right | \! P, \,  T \,
\rangle _{\xi^+ =0, \, \vec\xi_T=0}
= \frac{M^2}{(P^+)^2} \, S_{LL} \, f_{3LL} (x), 
\nonumber \\
\! \!
\Phi^{[\mathbf{1}]} (x,P,T) &
= \! \int \!  \frac{d\xi^-}{4\pi} \, e^{ixP^+ \xi^-}
\langle \, P , T \left | \, 
\bar\psi (0) 
\, \psi (\xi)  \, \right | \! P, \,  T \,
\rangle _{\xi^+ =0, \, \vec\xi_T=0}
 = \frac{M}{P^+} \, S_{LL} \, e_{LL} (x) ,
\label{eqn:correlation-pdfs}
\end{align}
where $\alpha$ is the transverse index $\alpha=1$ or 2 and 
this transverse $\alpha$ notation is used throughout 
this paper.
The tensor-polarized PDFs $f_{1LL}(x)$, $f_{LT}(x)$, $f_{3LL}(x)$, 
and $e_{LL}(x)$ are expressed by the matrix elements of the
nonlocal operators with different $\gamma$ matrices
in Eq.\,(\ref{eqn:correlation-pdfs}).
The collinear PDFs are often written from the TMDs as
\begin{align}
f (x) = \int d^2 k_T f (x, k_T^{\, 2}) .
\label{eqn:collinear-pdfs}
\end{align}
However, we should note that this integral equation could not be
a general relation due to the ultraviolet divergence
in the transverse-momentum integral \cite{rogers-2020}.
Futhermore, the TMDs $f^{\,\prime}_{LT} (x, k_T^{\, 2})$ and 
$f^{\perp}_{LT} (x, k^{\, 2}_T)$ are combined to express them as
\cite{ks-tmd-2021}
$ f_{LT} (x, k_T^{\, 2}) \equiv f^{\,\prime}_{LT} (x, k_T^{\, 2})
 - \frac{k_T^{\, 2}} {2M^2} \, f^{\perp}_{LT} (x, k^{\, 2}_T)$.
Therefore, two functions among $f_{LT}$, $f^{\,\prime}_{LT}$, and
$f^{\perp}_{LT}$ are independent, and $f_{LT}$ and $f^{\perp}_{LT}$
are selected in the table IV of Ref.\,\cite{ks-tmd-2021}.
In Eq.\,(\ref{eqn:correlation-pdfs}), $M$ is the spin-1 hadron mass,
$f_{1LL}$, $f_{LT}$, and $f_{4LL}$ are twist-2, twist-3,
and twist-4 distribution functions, respectively.
In this way, the collinear correlation function is written up to twist-4 as
\begin{align}
\Phi (x,P,T) 
= \frac{1}{2} \bigg [
S_{LL} \, \slashed{\bar n} \, f_{1LL} (x) 
+ \frac{M}{P^+} \, S_{LL} \, e_{LL} (x) 
+ \frac{M}{P^+} \, \slashed{S}_{LT} \, f_{LT} (x) 
+ \frac{M^2}{(P^+)^2} \, S_{LL} \, \slashed{n} \, f_{3LL} (x) 
\bigg ] .
\label{eqn:collinear-correlation-pdfs}
\end{align}
\ \vspace{-0.60cm}

\noindent
Then, the matrix element of Eq.\,(\ref{eqn:trace-correlation})
is expressed by the Fourier transform of 
$\Phi^{[\gamma^\mu]} (x,P,T)$ expressed by the PDFs as
\begin{align}
& \! \! \! \! \! \! 
\langle \, P , T \left | \, \bar\psi (0) 
\, \gamma^\mu \, 
\psi (\xi)  \, \right | \! P, \,  T \,
\rangle _{\xi^+ =0, \, \vec\xi_T=0} 
\nonumber \\
& \! \! \! \! \! \! \!
= \! 
\int_{-1}^1 dx e^{-ixP^+ \xi^-}
2 P^+  \left [ S_{LL} \, \bar n^\mu \, f_{1LL} (x) 
+ \frac{M}{P^+} \, S_{LT}^{\, \mu} \, f_{LT} (x) 
+ \frac{M^2}{(P^+)^2} \, 
S_{LL} \, n^\mu \, f_{3LL} (x)  
   \right ] .
\label{eqn:vector-matrix}
\end{align}
Therefore, the matrix element of the vector operator 
is given by the three collinear PDFs $ f_{1LL} (x)$, $f_{LT} (x)$,
and $f_{3LL} (x)$ for the tensor-polarized hadron.
In particular, we derive a useful relation
and a sum rule for the twist-3 functions $f_{LT} (x)$ in this work.

Since antiquark distribution are discussed later in deriving
twist-2 relations, we briefly explain them.
We define the collinear antiquark correlation function 
in the same way as \cite{mulders-2014-lect,ks-trans-g-2020}
\begin{align}
\bar \Phi_{ij} (x, P, T ) 
& = - \int  \frac{d\xi^-}{2\pi} \, e^{-ixP^+ \xi^-}
\langle \, P , T \left | \, 
\bar\psi _j (0)
\,  W (0, \xi \, |n)  \,
\psi _i (\xi) \, \right | \! P, \,  T \,
\rangle _{\xi^+ =0, \, \vec\xi_T=0}
\nonumber \\
& = - \Phi_{ij} (-x, P, T ) .
\label{eqn:correlation-pdf-anti}
\end{align}
Therefore, the antiquark correlation function is 
related to the quark correlation function at negative $x$,
so that the antiquark distributions are described by the quark 
distributions at negative $x$, $q(x<0)$.
On the other hand, the charge-conjugate correlation function,
in which the antiquark distributions $\bar q(x)$ are defined, is 
given by the conjugate spinor $\psi^C \equiv C \bar\psi^T$
with $C=i \gamma^2 \gamma^0$ and $A_\mu^C = - A_\mu$ as
\begin{align}
\Phi_{ij}^C (x, P, T ) 
= \int  \frac{d\xi^-}{2\pi} \, e^{ixP^+ \xi^-}
\langle \, P , T \left | \, 
\bar \psi_j^C (0)
\, W^C (0, \xi \, |n) \,
\psi _i^C (\xi) \, \right | \! P, \,  T \,
\rangle _{\xi^+ =0, \, \vec\xi_T=0} .
\label{eqn:correlation-pdf-anti-c}
\end{align}
These equations indicate the relation between them as
\begin{align}
\Phi^C (x, P, T ) = - C \, [ \bar\Phi (x, P, T ) ]^{\cal T} C^\dagger ,
\label{eqn:charge-conj-corr}
\end{align}
where ${\cal T}$ indicates the transposed matrix.
The antiquark distributions $\bar q (x)$ are then 
defined by using this conjugate correlation function.
The relations between the various antiquark distributions
$\bar q (x)$ and the corresponding ``quark" distributions $q (x<0)$
are explicitly written in the end of Sec.\,\ref{collinear-pdfs}.
Next, we show relations between the antiquark distributions
and the quark ones at negative $x$.
The relation between the correlation function $\bar \Phi$
and the conjugate one $\Phi^C$ is
\begin{align}
\Phi^{C[\Gamma]} (x, P, T ) 
= \left \{
\begin{array}{l}
   +\bar \Phi^{[\Gamma]} (x, P, T ) 
=  -     \Phi^{[\Gamma]} (-x, P, T )
 \ \  \text{for } \Gamma=\gamma^\mu,\ \sigma^{\mu\nu},\ i\gamma_5 \sigma^{\mu\nu}
\\
   -\bar \Phi^{[\Gamma]} (x, P, T ) 
=  +     \Phi^{[\Gamma]} (-x, P, T )
 \ \  \text{for } \Gamma=\mathbf{1},\ i\gamma_5,\ \gamma_5 \gamma^\mu
\end{array}
\right. ,
\label{eqn:correlation-pdf-anti-relations}
\end{align}
from Eqs.\,(\ref{eqn:correlation-pdf-anti}), (\ref{eqn:correlation-pdf-anti-c}), 
(\ref{eqn:charge-conj-corr}), and (\ref{eqn:trace-correlation}).
By taking $\Gamma=\gamma^\mu$ or $\mathbf{1}$
in Eqs.\,(\ref{eqn:collinear-correlation-pdfs}) 
and (\ref{eqn:correlation-pdf-anti-relations})
the antiquark distributions are related to the quark distributions as
\begin{align}
\bar f_{1LL} (x) & = - f_{1LL} (-x) , \ \ 
\bar  f_{LT} (x)   = - f_{LT}  (-x) , \ \ 
\bar f_{3LL} (x) = - f_{3LL} (-x) , 
\nonumber \\
\bar e_{LL} (x) & = e_{LL} (-x) ,
\label{eqn:antiquark-spin1}
\end{align}
where the sign is opposite for the chiral-odd 
distribution function $e_{LL} (x)$.


Let us consider the matrix element of non-local vector operator
$ \bar\psi(0) \gamma^{\mu} \psi(\xi) $ in the region which is not
necessarily on the lightcone, where Eq.\,(\ref{eqn:vector-matrix}) was obtained,
for calculating the matrix element with its derivative.
The Fourier transform of the vector-operator matrix element 
is the correlation function, which is expanded by the linear terms 
of the tensor polarization $T^{\mu\nu}$ as given in
Eq.\,(20) of Ref.\,\cite{ks-tmd-2021}.
Therefore, the matrix element should be generally expressed
in terms of three terms 
$(\xi \cdot T \cdot \xi) P^\mu$,
$(\xi \cdot T \cdot \xi) \xi^\mu$, and $T^{\mu\nu} \xi_\nu$
which are linear in the tensor polarization 
with the available Lorentz vectors $P^\mu$, $\xi^\mu$,
and $T^{\mu\nu} \xi_\nu$ as
\begin{align}
& \langle \, P ,  T \left | \,  \bar\psi(0)  \gamma^{\mu}  \psi(\xi) 
  \, \right | P \,  , T \, \rangle
\nonumber \\ 
& \ \hspace{1.0cm}
= \int_{-1}^1 dx \, e^{-i x P\cdot \xi}
  \left[ \,  \xi \cdot T \cdot \xi  \left \{  A(x) \, P^\mu
          + B(x) \, \xi^\mu \right \}
          + C(x) \, T^{\mu\nu} \xi_\nu
  \, \right]  ,
\label{eqn:vector-matrix-1}
\end{align}
where $\xi \cdot T \cdot \xi$ is defined by
$\xi \cdot T \cdot \xi = \xi_\mu T^{\mu\nu} \xi_\nu$
and $\xi$ may not be on the lightcone.
The term $T^{\mu\nu} P_\nu$ does not exist
in Eq.\,(\ref{eqn:vector-matrix-1})
because it vanishes identically $T^{\mu\nu} P_\nu=0$.
Here, $A(x)$, $B(x)$, and $C(x)$ are coefficients to be determined.
In this expansion, twist-4 effects are neglected, 
so that the twist-4 function $f_{3LL} (x)$ 
does not appear in the following discussions.
The tensor polarization $T^{\mu\nu}$ contains
the three types of polarizations 
$S_{LL}$, $S_{LT}^\mu$, and $S_{TT}^{\mu\nu}$,
and the right-hand side of Eq.\,(\ref{eqn:vector-matrix-1})
should be expressed by them.
We find the factors $A(x)$, $B(x)$, and $C(x)$
so that Eq.\,(\ref{eqn:vector-matrix-1}) becomes
Eq.\,(\ref{eqn:vector-matrix}) in the lightcone limit
$\xi^2 \to 0$ ($\xi^\mu = \xi^- n^\mu$, $\xi^+ = \vec \xi_T = 0$).
In this limit, the factors 
$\xi \cdot T \cdot \xi$ and $T^{\mu\nu} \xi_\nu$ are
expressed by the tensor polarization factors
$S_{LL}$ and $S_{LT}^\mu$  
by using Eq.\,(\ref{eqn:spin-1-tensor-1}) as 
\begin{align}
\xi \cdot T \cdot \xi
& = \frac{2}{3 \, M^2} (P^+ \xi^-)^2 S_{LL} ,
\nonumber \\
T^{\mu\nu} \xi_\nu  
& = \frac{2}{3 \, M^2} 
   \left [ (P^+)^2 \xi^- \bar n^\mu - \frac{1}{2} M^2 \xi^- n^\mu  \right ] S_{LL} 
   + \frac{1}{2 \, M} P^+ \xi^- S_{LT}^\mu .
\label{eqn:zTz-Tz}
\end{align} 
Then, $A(x)$, $B(x)$, and $C(x)$ are obtained  as
\begin{align}
A (x) & = \frac{3 \, M^2}{(P \cdot \xi)^2} \left [ f_{1LL} (x) 
               - \frac{4}{3} f_{LT} (x) \right ], \ 
B (x)  = \frac{3 \, M^4}{2(P \cdot \xi)^3}  \left [ 
               - f_{1LL} (x) + \frac{8}{3} f_{LT} (x)
               \right ] , 
\nonumber \\
C (x) & = \frac{4 \, M^2}{P \cdot \xi} f_{LT} (x)  ,
\label{eqn:ABC}
\end{align} 
in the lightcone limit with $P \cdot \xi = P^+ \xi^-$.
In order to derive a twist-2 relation and a sum rule, 
we need to investigate the twist-3 matrix element
$ \xi_\mu \langle \, P ,  T \big | \,  \bar\psi(0) 
(\partial^\mu \gamma^\alpha  - \partial^\alpha \gamma^\mu )
\psi(\xi)   \,  \big | P \,  , T \, \rangle $
as the derivative operator is given 
in Eqs.\,(\ref{eqn:twist-3-derivative}) and (\ref{eqn:derivative-op})
for $g_2$.
Taking the transverse index $\alpha=1$ or 2
and considering the lightcone limit,
we obtain the relation 
\begin{align}
& 
\xi_\mu \langle \, P ,  T \left | \,  \bar\psi(0) 
(\partial^\mu \gamma^\alpha  - \partial^\alpha \gamma^\mu )
\psi(\xi)   \, \right | P  , T \, \rangle 
\nonumber \\
&
=  2 M S_{LT}^{\, \alpha} \int_{-1}^1 
dx \, e^{-i x P^+ \xi^-} 
\left [ - \frac{3}{2} \, f_{1LL}(x)
+ f_{LT}(x) - \frac{d}{dx} \left\{  x f_{LT}(x) \right\} \right ] ,
\label{eqn:matrix-derivative}
\end{align}
from Eq.\,(\ref{eqn:vector-matrix-1}).
In this way, it becomes possible to identify the twist-3 part
of the function $f_{LT}(x)$ in connection with
the twist-2 function $f_{1LL}(x)$ .

Since the nonlocal operator of the left-hand side 
in Eq.\,(\ref{eqn:matrix-derivative}) gives rise to 
the twist-3 operators as explained in Sec.\,\ref{ww-bc-sum},
the right-hand side should vanish if higher-twist effects are
neglected. It leads to the WW- and BC-like twist-2 relations
for $f_{LT}$ and $f_{1LL}$.
However, we investigate further in this work by defining possible 
multiparton distribution functions for the tensor-polarized spin-1 hadron,
and then we explicitly show that the left-hand side is expressed
by these multiparton distribution functions.
Namely, we try to obtain the full decomposition of $f_{LT}$
into the twist-2 and twist-3 terms.


\subsection{Twist-3 matrix element and multiparton distribution functions}
\label{twist-3-multiparton}

For specifying twist-3 effects, we derived the expression
of twist-3 terms in Eq.\,(\ref{eqn:matrix-derivative})
in terms of the tensor-polarized distribution functions.
In general, the twist-3 terms are described by 
multiparton (three-parton in this work) distribution functions
\cite{kt-1999,bb-1988,Ball:1998ff}.
In order to derive a twist-2 relation and a sum rule
for the tensor-polarized PDFs,
we try to connect the derivative terms 
in the left-hand side of Eq.\,(\ref{eqn:matrix-derivative}) to 
the multiparton distribution functions.
For this purpose, we try to express the derivative terms
by the nonlocal quark-gluon operators 
in this subsection.

The Fock-Schwinger gauge $x_{\mu} A^{\mu}(x)=0$ is used 
in our formalism, so that the gluon field is expressed by 
the field strength tensor by introducing the variable $t$ as
\begin{align}
A^{\nu}(\xi)=\int^1_0 dt \, t \, \xi_{\mu} G^{\mu \nu}(t\xi),
 \label{eqn:fs1-1}
\end{align}
where $G^{\mu \nu}= \partial^\mu A^\nu - \partial^\nu A^\mu
                  - ig\left[A^{\mu}, A^{\nu} \right]  $.
The gauge link is generally expressed as
\begin{align}
W (0,\xi)
=  \pazocal{P} \exp \left[ - ig \int^1_0 \! dt \, \xi_\mu
A^{\mu} (t \xi)  \right],
\label{eqn:gl1}
\end{align}
where the integral path is the direct one from $0$ to $\xi$.
If the Fock-Schwinger gauge is taken, the gauge link becomes unity. 
However, the total derivative $\bar\partial^\alpha W$
does not vanish and it is used for relating the derivative relation 
of Eq.\,(\ref{eqn:matrix-derivative}) to the field tensor
and subsequently to the multiparton distribution functions.
We consider that the total derivative 
$\left. \frac{\partial}{\partial (\Delta \xi^\rho)} W (\Delta \xi, \xi+\Delta \xi ) 
 \right|_{\Delta \xi \rightarrow 0}$,
which is given by the field tensor $G^{\rho\mu}$ and
the gluon field $A^\rho$ as
\begin{align}
&
\bar\partial_\rho W (0, \xi) \equiv
\left. \frac{\partial}{\partial (\Delta \xi^\rho)} W (\Delta \xi, \xi+\Delta \xi) 
 \right|_{\Delta \xi \rightarrow 0}
= -ig \int^1_0 dt \, \xi^{\nu} G_{\rho \nu}(t\xi) 
        -ig \left[A_{\rho}(\xi)- A_{\rho}(0)\right] .
\label{eqn:gl3}
\end{align}
From Eq.\,(\ref{eqn:gl3}), the derivative of the local operator 
$\bar{\psi}(0) W \Gamma \psi (\xi)$ becomes
\begin{align}
\! 
\bar\partial_\rho \bar\psi (0)  &  W (0, \xi) \Gamma \psi (\xi) 
= \bar{\psi}(0) \left( \overleftarrow{D}_{\rho} 
        + \overrightarrow{D}_{\rho}  \right)  \Gamma \psi (\xi) 
-ig \int^1_0 dt \, 
\xi^{\nu}   G_{\rho \nu}(t\xi)
 \bar{\psi}(0) \Gamma \psi (\xi) ,
\label{eqn:gl4-1}
\end{align}
where the covariant derivatives are given by
$\overleftarrow{D}_{\rho}=\overleftarrow{\partial}_{\rho}+igA_{\rho}$
and
$\overrightarrow{D}_{\rho}=\overrightarrow{\partial}_{\rho}-igA_{\rho}$.

Next, we express the left-hand side of Eq.\,(\ref{eqn:matrix-derivative}) 
in terms of the covariant derivatives and the field tensor as
\begin{align}
\bar{\psi}(0)  (\partial^{\mu} \gamma^{\alpha}  
   - \partial^{\alpha} \gamma^{\mu} ) \psi (\xi)  
& = \bar{\psi}(0)  ( \overrightarrow{D}^{\mu} \gamma^{\alpha}  
 - \overrightarrow{D}^{\alpha} \gamma^{\mu} ) \psi (\xi)  
\nonumber \\
& 
 - \bar{\psi}(0) \gamma^{\mu} \psi (\xi) \,
       ig \int^1_0 dt \, t \, \xi_{\rho} G^{\rho \alpha}(t \xi) .
 \label{eqn:gl5-1}
\end{align}
The index $\alpha$ is the transverse one ($\alpha=1,\,2$);
however, all the following equations within this subsection are valid
as a general 4-dimensional Lorentz index.
By the identity 
\begin{align}
 \gamma^{\rho}\sigma^{\alpha \mu}
 = i(g^{\alpha \rho} \gamma^{\mu} -g^{\mu \rho} \gamma^{\alpha})
  -\epsilon^{\alpha \mu \rho \sigma} \gamma_{\sigma} \gamma_5 ,
\label{eqn:gammat1-1}
\end{align}
the covariant derivative term becomes
$\vec D^{\mu} \gamma^{\alpha}  - \vec D^{\alpha} \gamma^{\mu} 
= i (\overrightarrow{\slashed{D}} \sigma^{\alpha\mu} 
   - \sigma^{\alpha\mu} \overrightarrow{\slashed{D}}) /2$.
Then, using Eq.\,(\ref{eqn:gl4-1}) with the $\Gamma$ factor 
$\Gamma = \gamma^\rho \sigma^{\alpha\mu}$,
we obtain the first derivative terms in the right-hand side 
of Eq.\,(\ref{eqn:gl5-1}) as
\begin{align}
\bar{\psi}(0)  \big ( \overrightarrow{D}^{\mu}  \gamma^{\alpha} 
 &  - \overrightarrow{D}^{\alpha} \gamma^{\mu}  \big ) \psi (\xi)  
= -\frac{i}{2}  \bar{\psi}(0)
      \big ( \sigma^{\alpha \mu} \overrightarrow{\slashed{D}}
            + \overleftarrow{\slashed{D}} \sigma^{\alpha \mu} \big )
       \psi (\xi)  
\nonumber \\
& \ \ \ 
 -\frac{g}{2}  \int^1_0 dt \xi_{\nu} G^{\rho \nu}(t \xi) \bar{\psi}(0) 
   \gamma_{\rho} \sigma^{\alpha \mu} \psi(\xi)
 +\frac{i}{2}  \bar{\partial}_{\rho} \{ \bar{\psi}(0) \gamma^{\rho}
    \sigma^{\alpha \mu} \psi (\xi) \} .
  \label{eqn:gl8-1}
\end{align}
The third term of the right-hand side ($\gamma_\rho \sigma^{\alpha\mu}$)
is written by the antisymmetric tensor $\epsilon^{\alpha\mu\rho\sigma}$
of Eq.\,(\ref{eqn:gammat1-1}), and then it is
given by the dual field tensor $\widetilde{G}^{\mu\nu}$ as
\begin{align}
\xi_{\mu} \xi_{\nu}G_{\rho \tau}(t \xi)  
g^{\nu \tau} \epsilon^{\alpha \mu \rho \sigma}
= 2 \xi_{\mu} \xi^{\alpha} \widetilde{G}^{\mu \sigma}
- 2 \xi^2 \widetilde{G}^{\alpha \sigma} 
  +\xi_{\mu} \xi^{\tau}G_{\tau \rho} \epsilon^{\alpha \mu \rho \sigma }
  +2 \xi_{\mu} \xi^{\sigma} \widetilde{G}^{\alpha \mu} ,
\label{eqn:gl10-1}
\end{align}
where the relation 
$ 
g^{\nu \tau} \epsilon^{\alpha \mu \rho \sigma}
= g^{\nu \alpha} \epsilon^{\tau \mu \rho \sigma}
+g^{\nu \mu} \epsilon^{\alpha \tau \rho \sigma}
+g^{\nu \rho} \epsilon^{\alpha \mu \tau \sigma}
+g^{\nu \sigma} \epsilon^{\alpha \mu \rho \tau} 
$  
was used.
We may note that the third term in the right-hand side 
of Eq.\,(\ref{eqn:gl10-1}) is identical to the left-hand side
with the minus sign, so that the factor of 2 in front of
the dual field tensor is dropped for calculating
the left-hand-side term.
Therefore, the third term of Eq.\,(\ref{eqn:gl8-1}) 
contracted with $\xi_\mu$ becomes
\begin{align}
 \! \! \! \! \! 
-\frac{g}{2} \xi_{\mu} \int^1_0 \! dt \, \xi^{\nu} 
   G_{\rho \nu}(t \xi) \bar{\psi}(0)
   \gamma^{\rho} \sigma^{\alpha \mu} \psi(\xi) 
& = \frac{g}{2}  \int^1_0 \! dt \, \bar{\psi}(0)  \bigg [  
-i \xi_{\mu} G^{\alpha \mu}(t \xi)  \slashed{\xi}
 + \xi_{\mu} \widetilde{G}^{\alpha \mu}(t \xi)  \slashed{\xi} \gamma_5
\nonumber \\
& \ \ \ 
-    \left \{ \xi^2 \widetilde{G}^{\alpha \sigma} (t \xi)
   -  \xi_{\mu} \xi^{\alpha}
   \widetilde{G}^{\mu \sigma} (t \xi) \right \}
    \gamma_{\sigma} \gamma_{5}  \, \bigg ] \psi (\xi) .
\label{eqn:gl11-1}
\end{align} 
Substituting Eqs.\,(\ref{eqn:gl8-1}) and (\ref{eqn:gl11-1}) 
into Eq.\,(\ref{eqn:gl5-1}), we obtain
\begin{align}
& \! \! \! \! \!
\xi_{\mu} \, \bar{\psi}(0) \, ( 
    \overrightarrow{ \partial}^{\mu} \gamma^{\alpha}
   - \overrightarrow{\partial}^{\alpha} \gamma^{\mu} ) \, \psi (\xi) 
  = \! g  \! \int^1_0 \! dt \, \bar{\psi}(0) \! \left[  
i(t-\frac{1}{2}) G^{\alpha \mu}(t \xi) 
- \frac{1}{2} \gamma_5 \widetilde{G}^{\alpha \mu}(t \xi)  
 \right] \! 
 \xi_{\mu} \slashed{\xi} \psi (\xi)
\nonumber \\
& \ \hspace{-0.67cm}
 +\frac{g}{2}  \int^1_0 dt \, \bar{\psi}(0) 
   \left[  
 \xi_{\mu} \xi^{\alpha} \widetilde{G}^{\mu \sigma}(t\xi) 
 -\xi^2 \widetilde{G}^{\alpha \sigma}(t\xi)
 \right] \! \gamma_{\sigma} \gamma_{5} 
  \psi (\xi)
\nonumber \\ 
& \ \hspace{-0.67cm}
 -\frac{i}{2} \xi_{\mu}  \bar{\psi}(0)\sigma^{\alpha \mu}
      ( \overrightarrow{\slashed{D}} -m_q ) \psi (\xi)  
 -\frac{i}{2} \xi_{\mu} \bar{\psi}(0) 
      ( \overleftarrow{\slashed{D}} + m_q) 
     \sigma^{\alpha \mu} \psi(\xi) 
 +\frac{i}{2} \xi_{\mu}  \bar{\partial}_{\rho} 
  \{ \bar{\psi}(0) \gamma^{\rho} \sigma^{\alpha \mu} \psi(\xi) \} ,
  \label{eqn:gl12-1}
\end{align}
where the quark-mass term like the second line of Eq.\,(\ref{eqn:derivative-op})
does not exist due to the operator difference.
In this equation, the first term in the second line of RHS
is a twist-4 contribution 
and the next term vanishes in the lightcone limit $\xi^2=0$. 
The first two terms in the third line are 
the equation-of-motion terms, 
and the last term could be neglected since the total derivative term
vanishes in the forward matrix elements \cite{bb-1988,kt-1999}.
Therefore, at the twist-3 level, the relation is given by
\begin{align}
\xi_{\mu} \, \bar{\psi}(0) \big (
    \partial^{\mu} \gamma^{\alpha} 
   -\partial^{\alpha} \gamma^{\mu} \big ) \psi (\xi) 
= g  \int^1_0 dt \, \bar{\psi}(0) \!  \bigg [  
\, i \left ( t-\frac{1}{2} \right ) G^{\alpha \mu}(t\xi) 
- \frac{1}{2} \gamma_5 \widetilde{G}^{\alpha \mu}(t\xi)    \, \bigg] 
\xi_{\mu} \slashed{\xi} \,  \psi (\xi) ,
\label{eqn:gl12-mod1}
\end{align}
In this way, the derivative terms 
in the left-hand side of Eq.\,(\ref{eqn:matrix-derivative})
are given by the field tensor $G^{\mu\nu}$
and the dual one $\tilde G^{\mu\nu}$.


The next step is to relate the field-tensor terms of
Eq.\,(\ref{eqn:gl12-mod1}) to twist-3 multiparton 
(three-parton) distribution functions.
For this purpose, we define a quark-gluon-quark 
correlation function $\Phi_G^\mu (x_1, x_2)$ 
for a tensor-polarized spin-1 hadron
in terms of the field tensor $G^{\mu \nu}$.
For defining the correlation function, only the transverse component 
of the gluon field is considered in the lightcone formalism
as the leading term \cite{jaffe-1996,brodsky-1998,mulders-2014-lect}.
We have been using the Fock-Schwinger gauge ($\xi \cdot A=0$) 
in this paper; however, it is identical to the lightcone gauge ($n \cdot A=0$)
in the lightcone limit because of the relation 
$\xi \cdot A = \xi^- n \cdot A = \xi^- A^+ =0$.
The multiparton (quark-gluon-quark) correlation function is defined
by using the transverse gluon field $A^\alpha (= A_T^\alpha)$,
where $\alpha$ is taken as the transverse index $\alpha=1$ or 2.
In the lightcone gauge $A^+=0$,
the field tensor is expressed by the gluon field
as $G^{+\alpha} = \partial^+ A^\alpha$,
so that the correlation function is defined with $G^{+\alpha}$ as
\begin{align}
(\Phi_G^\alpha)_{ij} (x_1, x_2)=
\int  \! \frac{d \xi_1^-}{2\pi}   \frac{d \xi_2^-}{2\pi}  
 \,    e^{i x_1 P^+ \xi_1^-}  e^{i (x_2-x_1) P^+ \xi_2^-} 
\langle \, P, T \left | \,  \bar\psi _j (0) \, 
g \, G^{+ \alpha}( \xi_2^- ) \, \psi _i (\xi_1^-)  \,
  \right | P, T \, \rangle .
\label{eqn:3predef-1}
\end{align} 
Due to the relation $G^{+\alpha} = \partial^+ A^\alpha$,
the same correlation-function expression, which is
derived in the following,
should be valid for the correlation function defined
with the gluon field $A^\alpha$ instead of $G^{+ \alpha}$,
except for minor changes discussed after Eq.\,(\ref{eqn:time-rev-PG}).
Here, the gauge link is not written because of the lightcone gauge.
The conditions of Hermiticity, parity invariance, and time-reversal 
invariance are given in Refs.\cite{mulders-2014-lect,ks-tmd-2021}.

Next, we try to express the multiparton correlation function 
in terms of possible Lorentz vectors with the index $\alpha$.
There was some study on the multiparton
correlation function for the deuteron
in Ref.\,\cite{ma-spin-1-2013}.
In this work, we try to provide the full expression 
for the  multiparton correlation function for spin-1 hadrons,
and then we relate them to the twist-2 relation and sum rule
for $f_{LT}$.
Here, only the tensor-polarization is considered since
the unpolarized and vector-polarization correlation functions
have already investigated for the spin-1/2 nucleons.
Then, the correlation function of Eq.\,(\ref{eqn:3predef-1})
should be proportional to the tensor polarization 
$T^{\mu\nu}$ as shown in Eq.\,(20) of Ref.\,\cite{ks-tmd-2021}.
The correlation function in Eq.\,(\ref{eqn:3predef-1}) has
Dirac spinor indices $i$ and $j$, so that it is expressed 
by the Dirac $\gamma$ matrices
such as $\gamma^\mu$ and $\gamma^\mu\gamma^\nu$.
Three and more $\gamma$ types are not independent from these
one- and two-$\gamma$ types because of the relation
$\gamma^\mu \gamma^\rho \gamma^\nu 
 = S^{\mu\rho\nu\beta} \gamma_\beta 
 + i \epsilon^{\mu\rho\nu\beta} \gamma_\beta \gamma_5$ with
$S^{\mu\rho\nu\beta} 
 = g^{\mu\rho}g^{\nu\beta} + g^{\mu\beta}g^{\rho\nu} 
 - g^{\mu\nu}g^{\rho\beta}$.

In terms with $T^{\mu\alpha}$,
the Lorentz index $\mu$ should be contracted with other vectors.
We note that the hadron momentum $P^\mu$ is expressed 
by the two lightcone vectors $n^\mu$ and $\bar n^\mu$
($P^\mu = P^+ \bar n^\mu + M^2 n^\mu/(2P^+)$),
where the $n^\mu$ term is suppressed by the factor $O(M^2/(P^+)^2)$.
First, let us take the contraction of $T^{\mu\alpha}$ in
Eq.\,(\ref{eqn:spin-1-tensor-1}) with the lightcone vector $n_\mu$ 
to obtain
$n_\mu T^{\mu\alpha} \slashed{\bar n} 
= \frac{P^+}{2M} S_{LT}^{\,\alpha} \slashed{\bar n}$,
where $\slashed{\bar n}$ is multiplied for including the $\gamma$ 
matrix, and the higher-twist $\slashed{n}$ terms are not included.
Namely, the term $n_\mu T^{\mu\alpha}$ is given 
by the tensor polarization vector $S_{LT}^{\,\alpha}$.
The term $\bar n_\mu T^{\mu\alpha}$ is expressed as
$\bar n_\mu T^{\mu\alpha} = - \frac{M^2}{2 (P^+)^2} n_\mu T^{\mu\alpha}$,
so that it is not an independent term.
In the same way, the term $\gamma_\mu T^{\mu\alpha}$ is not
independent from $n_\mu T^{\mu\alpha}$ as they are related
with each other by 
$n_\mu T^{\mu\alpha} \slashed{\bar n} = \gamma_\mu  T^{\mu\alpha}$
in the lightcone limit.
On the other hand, $\gamma_{T \mu} T^{\mu\alpha}  \slashed{\bar n}$ 
is an independent term.
Therefore, we have two independent terms associated with $T^{\mu\alpha}$:
\begin{align}
n_\mu T^{\mu\alpha} \slashed{\bar n} 
= \frac{P^+}{2M} S_{LT}^{\,\alpha} \slashed{\bar n},
\ \ 
\gamma_{T \mu} T^{\mu\alpha} 
\slashed{\bar n}
= \left ( \frac{1}{3} S_{LL} \gamma_T^\alpha 
           + \frac{1}{2} \gamma_{T\mu} S_{TT}^{\mu\alpha} \right )
\slashed{\bar n} .
\label{eqn:Tmu-alpha}
\end{align} 
As for independent $T^{\mu\nu}$ terms
where the index $\alpha$ comes from other vectors,
we have two possibilities
by noting the transverse components $\epsilon_{T}^{\alpha\nu}$
($\epsilon_T^{11}=-\epsilon_T^{22}=1$)
and $\gamma_T^\alpha$ as
\begin{align}
n_\mu T^{\mu\nu} \epsilon_{T\nu}^{\, \alpha} 
i \gamma_5 \slashed{\bar n}
= \frac{P^+}{2M} \epsilon_T^{\alpha\mu} S_{LT\mu}
i \gamma_5 \slashed{\bar n} 
, \ \ 
n_\mu T^{\mu\nu} n_\nu 
\gamma_T^\alpha \slashed{\bar n}
= \frac{2}{3} S_{LL} \frac{(P^+)^2}{M^2} \gamma_T^\alpha
\slashed{\bar n} . 
\label{eqn:other-terms}
\end{align} 
From these considerations,
the correlation function $\Phi_G^\alpha (x_1, x_2)$ of 
Eq.\,(\ref{eqn:3predef-1}) is generally expressed 
by four terms with the tensor polarizations as
\begin{align}
\Phi_G^\alpha (x_1, x_2) = \frac{M}{2} \bigg [  \, &
i S_{LT}^\alpha  \, F_{G,LT}(x_1, x_2)
- \epsilon_{T}^{\alpha \mu} S_{LT \mu} 
\gamma_5  G_{G,LT}(x_1, x_2)  
\nonumber \\
&
+ i S_{LL} \gamma^{\alpha} H_{G,LL}^\perp (x_1, x_2) 
+ i S_{TT}^{\alpha \mu} \gamma_{\mu}  H_{G,TT}(x_1, x_2) \bigg ] 
\slashed{\bar n} .
\label{eqn:3predef1-1}
\end{align} 
All the terms in the right-hand side of Eq.\,(\ref{eqn:3predef1-1})
satisfy the parity and time-reversal invariances. 
The functions 
$F_{G,LT}(x_1, x_2)$,
$G_{G,LT}(x_1, x_2)$,
$H_{G,LL}^\perp(x_1, x_2)$,
$H_{G,TT}(x_1, x_2)$
are twist-3 multiparton distribution functions.
The correlation function $\Phi_A^\alpha$ is defined by using
the gluon field $A^\alpha$ instead of the field thensor $G^{+\alpha}$
in Eq.\,(\ref{eqn:3predef-1}), and it is expressed by
real multiparton correlation functions. 
Therefore, the right-hand side of Eq.\,(\ref{eqn:3predef1-1})
contains the $i$ factor due to the derivative $\partial^+ (\to i P^+)$
in $G^{+\alpha}$.

We try to find properties of the multiparton distribution functions 
$F_{G,LT}(x_1, x_2)$,
$G_{G,LT}(x_1, x_2)$,
$H_{G,LL}^\perp(x_1, x_2)$,
$H_{G,TT}(x_1, x_2)$
under the exchange of variables $x_1$ and $x_2$ 
by using the Hermiticity condition.
For example, this Hermiticity relation is given for 
the first term of Eq.\,(\ref{eqn:3predef1-1}) as
$-i S_{LT}^\nu \slashed{\bar n}^\dagger 
F_{G,LT} (x_1, x_2)^*
= i S_{LT}^\nu \gamma^0 \slashed{\bar n} \gamma^0 F_{G,LT} (x_2, x_1)$.
For the real function of 
$F_{G,LT}(x_1, x_2)$, 
it becomes 
$F_{G,LT}(x_1, x_2)=-F_{G,LT}(x_2, x_1)$.
In this way, the Hermiticity condition is satisfied 
if the functions $F_{G,LT}$, $G_{G,LT}$, $H_{G,LL}^\perp$, and $H_{G,TT}$ 
are real and they have the properties 
\begin{alignat}{2}
F_{G,LT}(x_1, x_2) & = - F_{G,LT}(x_2, x_1), \ \ &
G_{G,LT}(x_1, x_2) & =G_{G,LT}(x_2, x_1),
\nonumber \\
H_{G,LL}^\perp(x_1, x_2) & =H_{G,LL}^\perp(x_2, x_1), \ \ &
H_{G,TT}(x_1, x_2) & =H_{G,TT}(x_2, x_1),  
\label{eqn:time-rev-PG}
\end{alignat} 
under the exchange of the variables $x_1$ and $x_2$.
If the multiparton distribution functions are defined
by taking the gluon field $A^\alpha$ instead of $G^{+\alpha}$ 
in Eq.\,(\ref{eqn:3predef-1}) and the functions 
($F_{A,LT}$, $G_{A,LT}$, $H_{A,LL}^\perp$, and $H_{A,TT}$)
are defined without the $i$ factors, 
these symmetric properties
have opposite signs, namely 
$F_{A,LT}(x_1, x_2) = F_{A,LT}(x_2, x_1)$, 
$G_{A,LT}(x_1, x_2) = -G_{A,LT}(x_2, x_1)$,
$H_{A,LL}^\perp(x_1, x_2) = -H_{A,LL}^\perp(x_2, x_1)$, and 
$H_{A,TT}(x_1, x_2) = -H_{A,TT}(x_2, x_1)$.

From Eqs.\,(\ref{eqn:3predef-1}) and (\ref{eqn:3predef1-1}),
the functions $F_{G,LT}$ and $G_{G,LT}$ are expressed 
as the correlation matrix-element forms as
\begin{align}
& 
S_{LT }^{\nu} F_{G,LT}(x_1, x_2)= 
- \frac{i}{2M}
\, g \!
\int \frac{d\xi_1^-}{2\pi} \frac{d\xi_2^-}{2\pi} 
\, e^{i x_1 P^+ \xi_1^-}  
\, e^{i (x_2-x_1) P^+ \xi_2^-} 
\nonumber \\
& 
\ \hspace{4.0cm}
\times
\langle \, P ,  T \left | \,  \bar\psi  (0) \,  
\slashed{n} n_\mu  
G^{\mu \nu}( \xi_2^- )  
\psi (\xi_1^-)  \, \right | P, T \, \rangle ,
\label{eqn:3predef2-1-G}
\\[-0.80cm]
\nonumber 
\end{align} 
and
\vspace{-0.10cm}
\begin{align}
& 
S_{LT }^{\nu} 
G_{G,LT}(x_1, x_2)= \frac{i}{2M} \, g \!
\int  \frac{d \xi_1^-}{2\pi} \frac{d \xi_2^-}{2\pi}   \,  
 e^{i x_1 P^+ \xi_1^-}  \,
 e^{i (x_2-x_1) P^+ \xi_2^-} 
\nonumber \\
& \ \hspace{4.0cm}
\times
\langle \, P ,  T \left | \, \bar\psi  (0) \,
i \gamma_5 \slashed{n} n_\mu
\tilde{G}^{\mu \nu}( \xi_2^- )   \psi (\xi_1^-)  
\, \right | P,   T \, \rangle .
\label{eqn:3predef2-1-F}
\end{align}
Equations (\ref{eqn:3predef2-1-G}) and (\ref{eqn:3predef2-1-F}) were 
obtained by the traces of 
$(\Phi_G^\alpha)_{ij} (x_1, x_2) (\slashed{n})_{ji}$ and
$(\Phi_G^\alpha)_{ij}\\ (x_1, x_2) (i \gamma_5 \slashed{n})_{ji}$,
respectively.
Then, the relations 
$\epsilon_T^{\alpha \mu} S_{LT \mu}
= g ^{\alpha 1} S_{LT}^2 - g^{\alpha 2} S_{LT}^1$ and
$n_\mu \tilde G^{\beta\gamma} \epsilon_{\beta\gamma}^{\ \ \mu\alpha}
\\ = g^{\alpha 1} 2 \tilde G^{+ 2} - g^{\alpha 2} 2 \tilde G^{+1}$
were used to reach to Eq.\,(\ref{eqn:3predef2-1-F}).
Defining the variable $t$ by $\xi_2^- = t \xi_1^-$ 
and calculating derivatives of $F_{G,LT}(x_1, x_2)$ and $G_{G,LT}(x_1, x_2)$ 
with respect to $x_1$ and $x_2$, we obtain the relation
with the field tensors $G^{\alpha\mu}$
and $\tilde G^{\alpha\mu}$ in Eq.\,(\ref{eqn:gl12-mod1}).
In this calculation, the principal integral
expressed by the sign function 
\begin{align}
\frac{i}{\pi} {\cal P} \int_{-\infty}^\infty 
d \omega \, \frac{1}{\omega} \, e^{- i \omega z}
= \epsilon (z)
= \left \{
\begin{array}{l}
+1 \ \  \text{for } z>0 
\\
-1 \ \  \text{for } z<0
\end{array}
\right. ,
\label{eqn:sign-fun}
\end{align}
is used. 
The integral region of $x_2$ is from $-1$ to $1$; however,
the integrand with the distribution functions vanish 
in the region $|x_2| \ge 1$, 
so that the integral region is extended 
to the one from $-\infty$ to $\infty$.
Here, ${\cal P}$ indicates the principal integral.
Using Eqs.\,(\ref{eqn:3predef2-1-G}) and (\ref{eqn:3predef2-1-F}),
we obtain the matrix element of the field tensors 
in the right-hand side of Eq.\,(\ref{eqn:gl12-mod1}) as
\begin{align}
& \! \! \! \!
\int  
\frac{d (P \cdot \xi)}{ 2 \pi} \, e^{i x_1 P \cdot \xi} \,
\bigg \langle \, P, T \, \bigg | \,
g \int^{1}_0   dt \, \bar{\psi}(0)   
\bigg [  
i \left ( t-\frac{1}{2} \right ) 
G^{\alpha\mu}(t\xi) 
- \frac{1}{2} \gamma_5   \,
\tilde G^{\alpha\mu}(t\xi) \bigg ]
\xi_\mu \slashed{\xi} \psi( \xi) 
\, \bigg | \, P, T \, \bigg \rangle _{\xi^+ = \vec \xi_T =0}   
\nonumber \\
& \! \! \!
= - 2M S_{LT }^{\nu}
{\cal P} \int_{-1}^1 dx_2 \frac{1}{x_1-x_2} 
\bigg [ \frac{\partial}{\partial x_1} 
   \left \{ F_{G,LT} (x_1, x_2)+G_{G,LT}(x_1, x_2) \right \}
\nonumber \\
& \ \hspace{4.5cm}
+ \frac{\partial}{\partial x_2} 
   \left \{ F_{G,LT} (x_2, x_1) +G_{G,LT}(x_2, x_1) \right \} \bigg ] .
\label{eqn:3pred4-1}
\end{align}
In this way, it becomes possible to separate the twist-3 effects expressed
by the multiparton distribution functions from the twist-2 ones, which is
essential in deriving the WW-like relation for the distribution
function $f_{LT}$.

The multiparton correction functions 
$\bar \Phi_G^{\alpha}$ and $\Phi_G^{\alpha, C}$ 
are defined for antiquarks in the same way with 
Eqs.\,(\ref{eqn:correlation-pdf-anti}),
(\ref{eqn:correlation-pdf-anti-c}), and
(\ref{eqn:charge-conj-corr})
by considering the definition for quarks in Eq.\,(\ref{eqn:3predef-1}).
The only extra factor is the gluon tensor $G^{+\alpha}$.
Noting the charge-conjugation property for the gluon field
$A_\mu^C = - A_\mu$, we find the relation which
has the opposite sign of Eq.\,(\ref{eqn:charge-conj-corr}) as
\begin{align}
\Phi_G^{\alpha, C} (x_1, x_2, P, T ) 
=  C \, [ \bar\Phi_G^{\alpha} (x_1, x_2, P, T ) ]^{\cal T} C^\dagger .
\label{eqn:charge-conj-corr-multi}
\end{align}
Therefore, the relations between the multiparton correlation functions 
for antiquarks have different signs from 
Eq.\,(\ref{eqn:correlation-pdf-anti-relations}) as
\begin{align}
\Phi_G^{\alpha,C[\Gamma]} (x_1,x_2) 
= \left \{
\begin{array}{l}   
   - \bar\Phi_G^{\alpha,[\Gamma]} (x_1,x_2) 
=  + \Phi_G^{\alpha,[\Gamma]} (-x_2,-x_1)
 \ \  \text{for } \Gamma=\gamma^\mu,\ \sigma^{\mu\nu},\ i\gamma_5 \sigma^{\mu\nu}
\\
   + \bar\Phi_G^{\alpha,[\Gamma]} (x_1,x_2)  
=  - \Phi_G^{\alpha,[\Gamma]} (-x_2,-x_1) 
 \ \  \text{for } \Gamma=1,\ i\gamma_5,\ \gamma_5 \gamma^\mu
\end{array}
\right. .
\label{eqn:correlation-3pdf-anti-relations}
\end{align}
Then, the multiparton distribution functions for antiquark are
obtained by noting these correlation-function relations as
\begin{alignat}{2}
\bar F_{G,LT} (x_1, x_2) & = F_{G,LT} (-x_2, -x_1), \ \ &
\bar G_{G,LT} (x_1, x_2) & = - G_{G,LT} (-x_2, -x_1),
\nonumber \\
\bar H_{G,LL}^\perp (x_1, x_2) & = H_{G,LL}^\perp (-x_2, -x_1), \ \ &
\bar H_{G,TT} (x_1, x_2) & 
=  H_{G,TT} (-x_2, -x_1) .
\label{eqn:3parton-antiquark}
\end{alignat} 
Therefore, the multiparton distribution functions for antiquarks are
described by the functions for quarks at negative $x_1$ and $x_2$;
however, they are also expressed by the functions 
$\bar F_{G,LT}$, $\bar G_{G,LT}$, $\bar H_{G,LL}^\perp$, and 
$\bar H_{G,TT}$.
Similar relations were given in the multiparton distribution functions
for the spin-1/2 nucleons in Ref.\,\cite{Kanazawa-2016}.

\subsection{\boldmath Twist-2 relation and sum rule}
\label{twist-2-relations}

After all these preparations on the tensor-polarized PDFs
of spin-1 hadrons, we are now ready to drive a twist-2 relation and 
a sum rule for the twist-3 distribution function $f_{LT}$.
First, the matrix element of the nonlocal operator
$\bar\psi(0) 
(\partial^\mu \gamma^\alpha 
 - \partial^\alpha \gamma^\mu ) 
\psi(\xi)$
with the transverse index $\alpha$
was expressed in terms of the tensor-polarized PDFs
$f_{1LL}(x)$ and $f_{LT}(x)$ in Eq.\, (\ref{eqn:matrix-derivative}).
Second, this nonlocal operator was expressed by the field tensor $G^{\alpha\mu}$
and its dual one $\tilde G^{\alpha\mu}$ in Eq.\,(\ref{eqn:gl12-mod1}).
Third, the matrix element of the field tensors was given by
the twist-3 multiparton distribution functions $G$ and $F$
in Eq.\,(\ref{eqn:3pred4-1}).
Combining Eqs.\, (\ref{eqn:matrix-derivative}), (\ref{eqn:gl12-mod1}), 
and (\ref{eqn:3pred4-1}), we obtain
\begin{align}
x \, \frac{df_{LT}(x)}{dx} = - 
\frac{3}{2} f_{1LL}(x)
- f_{LT}^{(HT)}(x),
\label{eqn:pdfredef4}
\end{align}
where 
the twist-3 multiparton-distribution part is defined as
\begin{align}
f_{LT}^{(HT)} (x)
= - {\cal P} \int_{-1}^1 dy \frac{1}{x-y} 
  \bigg[ \, & \frac{\partial}{\partial x} 
   \left \{ F_{G,LT} (x, y) 
+ G_{G,LT} (x, y) \right \}
\nonumber \\
&
   + \frac{\partial}{\partial y}  \left \{ F_{G,LT} (y, x) 
+ G_{G,LT} (y, x) \right \} \bigg] .
\label{eqn:fbar-twist-3}
\end{align}
Here, $HT$ indicates the higher twist.
Integrating Eq.\,(\ref{eqn:pdfredef4}) over $x$, we obtain
\begin{align}
f_{LT}(x)= \frac{3}{2} \int^{\epsilon (x)}_x \frac{dy}{y} f_{1LL}(y)
+\int^{\epsilon (x)}_x \frac{dy}{y} f_{LT}^{(HT)}(y),
\label{eqn:ww3}
\end{align}
by using the sign function of Eq.\,(\ref{eqn:sign-fun}).
Namely, the integral is from $x$ to $1$ if $x$ is positive
for the quark distribution,
and it is from $x$ to $-1$ if $x$ is negative
for the antiquark distribution.
In Eq.\,(\ref{eqn:ww3}), we obtained the full decomposition $f_{LT}$ 
into the twist-2 term and the twist-3 multiparton distribution
functions, as investigated for $g_2$ and $h_L$ in 
Refs.\,\cite{jj-1991,chiral-odd-twist-3,kt-1999}.

In Sec.\,\ref{ww-bc-sum}, the structure function $g_1$ is given
by the function $g_{1L}$ as $g_1 (x) = [ g_{1L} (x) + g_{1L} (-x) ]/2$
for describing both the quark and antiquark distributions in
the $x$ range of $0 \le x \le 1$. In the same way, we define
the distribution functions $f_{1LL}^+$, $f_{LT}^+$, and $f_{LT}^{(HT)+}$ as
\begin{align}
f^+ (x) \equiv f (x) + \bar f (x)
            =  f (x) -  f (-x),\ \ 
f=f_{1LL},\ f_{LT},\ f_{LT}^{(HT)},\ \ x>0.
\label{eqn:quark-antiquark}
\end{align}
Here, these functions are given for a single flavor.
From Eqs.\,(\ref{eqn:ww3}) and (\ref{eqn:quark-antiquark}),
we obtain a relation for $f_{LT}^+ (x)$ and 
$f_{1LL}^+ (x)$ 
as
\begin{align}
f_{LT}^+(x)= \frac{3}{2} \int^1_x \frac{dy}{y} \, f_{1LL}^+ (y)
+\int^1_x \frac{dy}{y} \, f_{LT}^{(HT)+}(y) .
\label{eqn:pdfredef5}
\end{align}
Since the function $f_{LT}^{(HT)+}(y)$ indicates the twist-3 effects
as given by the multiparton correlation functions,
this equation indicates that the twist-2 part of
the function $f_{LT}^+ (x)$ is expressed by the integral of $f_{1LL}^+ (x)$.
In this way, the twist-3 function $f_{LT}^+ (x)$ is expressed
by the twist-2 function and the remaining twist-3 one.
If the twist-3 part is neglected, the relation becomes
\begin{align}
f_{LT}^+(x)= \frac{3}{2} \int^1_x \frac{dy}{y} \, f_{1LL}^+(y) .
\label{eqn:flt-twist2}
\end{align}
This equation corresponds to the WW relation in Eq.\,(\ref{eqn:ww}).
It should be noted that the structure function $b_1$ 
or the tensor-polarized PDF $\delta_T q$ is given
by the $f_{1LL}$ as 
$-(3/2) f_{1LL}^+ 
= b_1^q + b_1^{\bar q}
= \delta_T q + \delta_T \bar q$ 
\cite{ks-tmd-2021}.
This new relation suggests that 
the tensor-polarized distribution function $f_{LT}(x)$ 
is expressed by the integral of $f_{1LL} (x)$ or
$b_1 (x)$ if higher-twist effects are ignored.
If the function $f_{2LT}(x)$ is defined by
\begin{align}
f_{2LT}(x) \equiv 
\frac{2}{3} f_{LT}(x) - f_{1LL}(x) ,
\label{eqn:f2ll}
\end{align}
Eq.\,(\ref{eqn:pdfredef5}) becomes
\begin{align}
f_{2LT}^+ (x) = -f_{1LL}^+ (x)+ \int^1_x \frac{dy}{y} \, f_{1LL}^+(y)
+ \frac{2}{3} \int^1_x \frac{dy}{y} \, f_{LT}^{(HT)} (y) .
\label{eqn:f2ll-2-3}
\end{align}
The distribution function $f_{2LL}^+(x)$ is expressed
by the twist-2 and twist-3 terms.
If the twist-3 term is neglected, we obtain a relation
\begin{align}
f_{2LT}^+ (x)=-f_{1LL}^+ (x)+ \int^1_x \frac{dy}{y} \, f_{1LL}^+(y) ,
\label{eqn:pdfredef7}
\end{align}
which is analogous to the WW relation for $g_1$ and $g_2$
in Eq.\,(\ref{eqn:ww}).
Furthermore, integrating this equation, we obtain
\begin{align}
\int_0^1 dx \, f_{2LT}^+ (x) =0 ,
 \label{eqn:pdfredef8}
\end{align}
which is analogous to the BC sum rule.
These relation and sum rule are useful in studying the tensor-polarized
distribution function $f_{2LT}^+(x)$ (or original $f_{LT}(x)$) 
as the WW relation and the BC sum rule provide strong constraints 
on determining the structure function $g_2 (x)$ for the nucleons.
Furthermore, considering the sum rule based on the parton model
$\int dx b_1 (x) = 0$ (or $\int dx f_{1LL}^+ (x) = 0$ by the notation
in this paper) \cite{b1-sum,Airapetian:2005cb}, 
which is valid if tensor-polarized antiquark distributions vanish,
we have the sum rule for the twist-3 function $f_{LT}$ itself as
\begin{align}
\int_0^1 dx \, f_{LT}^+ (x) =0 .
 \label{eqn:f_lt-sum}
\end{align}
Equations (\ref{eqn:pdfredef8}) and (\ref{eqn:f_lt-sum}) could be 
affected by the small-$x$ behavior of the distribution functions 
in the same way with the BC sum rule for $g_2$.
We proved these twist-2 relations in the tree level and have not
discussed perturbative QCD corrections.
At this stage, it is not obvious whether these relations are
satisfied in the structure-function level by
including coefficient functions as investigated
in $g_2$ \cite{Braun-2001}. We leave this issue for
a future project.

The tensor-polarized structure functions of spin-1 hadrons and nuclei
have been investigated since the end of 1980's. 
Due to lack of experimental measurements except for the HERMES experiment
for $b_1$, theoretical developments of this field were rather slow
in comparison with the spin physics of spin-1/2 nucleons.
However, we believe that bright future is ahead of us in the sense
that the tensor-polarized structure function $b_1$
and the gluon transversity, which are specific observables
in the spin-1 hadrons, will be measured at JLab in the middle of
2020's \cite{jlab-b1,jlab-gluon-trans} and 
such experiments will be proposed at Fermilab \cite{Fermilab-dy}
as the proton-deuteron Drell-Yan process.
The NICA facility will have the polarized-deuteron beam 
in the near future \cite{nica}, so that they could focus 
their studies on the structure functions of the spin-1 deuteron.
In addition, there are EIC projects in US and China
\cite{eic,eicc} to investigate the structure functions
of the spin-1 hadrons and nuclei in 2030's.
In the JLab measurements, the scale $Q^2$ is not large
in general, which enables to probe the twist-3
structure functions such as $g_2$. 
In the same way, higher-twist tensor-polarized
structure functions could become accessible at JLab
or future high-intensity accelerator facilities.
In this sense, our previous studies on general
twist-3 and twist-4 distribution functions \cite{ks-tmd-2021}
as well as this work should become useful in future.
In particular, the new twist-2 relation 
of Eq.\,(\ref{eqn:pdfredef7}) [or (\ref{eqn:flt-twist2})] 
and the sum rule of Eq.\,(\ref{eqn:pdfredef8}) 
[also Eq.\,(\ref{eqn:f_lt-sum}) in addition]
could become important for constraining the twist-3 
function $f_{2LT}(x)$ or $f_{LT}(x)$,
although the integral relations are always difficult 
to be tested due to the experimental inaccessibility at small $x$.
The tensor-polarized PDFs and structure functions will be measured 
at JLab, Fermilab, and NICA at relatively large $x$,
and the small-$x$ part will be investigated at EIC
and other high-energy lepton facilities, for example,
by a possible fixed target project of a linear or circular 
lepton collider.
According to the theoretical estimate on higher-twist 
tensor-polarized structure
functions in a few-GeV $Q^2$ region \cite{b1-convolution}, 
which is the typical kinematical region of JLab, 
they are not be much smaller than the leading-twist ones,
in the similar way to the 15--40\% breaking of 
the WW relation for $g_2$ \cite{accardi-2009}.
Therefore, the understanding of the higher-twist structure
functions are valuable also in determining the leading-twist functions
from actual measurements in future.
At this stage, the available information is very limited
for the tensor-polarized PDFs and structure functions. 
However, we hope to make progress on numerical studies of
them by considering the sum rules derived in this work.

\section{Summary}
\label{summary}

There are tensor-polarized PDFs and structure functions for spin-1 hadrons.
In this work, we derived a new useful twist-2 relation 
\begin{align}
f_{LT}(x)= \frac{3}{2} \int^{\epsilon (x)}_x dy \frac{f_{1LL}(y)}{y}
          +\int^{\epsilon (x)}_x dy \frac{f_{LT}^{(HT)}(y)}{y},
\nonumber 
\end{align}
where $\epsilon (x)=1$ ($-1$) at $x>0$ ($x<0$),
for the twist-3 distribution function $f_{LT}(x)$ 
and twist-2 one $f_{1LL}(x)$.
This equation indicates the quark and antiquark distributions at $x>0$ 
and $x<0$, respectively.
Defining the plus function by 
the quark and antiquark distributions as
$f^+ (x) = f(x) + \bar f(x)$ 
and neglecting the higher-twist term, this equation is written as
\begin{align}
f_{LT}^+(x)= \frac{3}{2} \int^1_x 
\frac{dy}{y} \, f_{1LL}^+(y) .
\end{align}
Namely, the twist-2 part of $f_{LT} (x)$ is expressed by
the integral of $f_{1LL}(x)$.
Since the integrand is given by the structure function 
$b_1^{q+\bar q} = - (3/2) f_{1LL}^+$, 
the twist-2 of $f_{LT}(x)$ is expressed
by the function $b_1$.
Using the function $f_{2LL}$ defined by 
$f_{2LL} = \frac{2}{3} f_{LT}-f_{1LL}$,
we obtained
\begin{align}
f_{2LT}^+ (x)=-f_{1LL}^+ (x)+ \int^1_x \frac{dy}{y} f_{1LL}^+ (y) .
\nonumber 
\end{align}
This relation is similar to the Wandzura-Wilczek relation for the polarized
structure functions $g_1$ and $g_2$ for the spin-1/2 nucleons.
It is useful in the sense that the twist-2 part
is constrained and the separation of higher-twist effects become clear. 
In addition, we showed that the sum rule
\begin{align}
& \int_0^1 dx \, f_{2LT}^+(x) =0 ,
\nonumber 
\end{align}
exists for $f_{2LL}$, and it constrains
the overall $x$-dependent functional form of $f_{2LL}$.
It is similar to the Burkhardt-Cottingham sum rule for $g_2$.
Furthermore, if the parton-model sum rule 
$\int dx f_{1LL}^+ (x) = 0$ 
($\int dx b_1^{q+\bar q} (x) = 0$) 
is applied in the case where the tensor-polarized antiquark
distributions vanish, it led to another sum rule
\begin{align}
& \int_0^1 dx \, f_{LT}^+(x) =0 .
\nonumber 
\end{align}
All these relations are valuable in investigating 
the tensor-polarized PDFs and structure functions in future.
For specifying twist-3 terms in deriving these relations,
we explained that four twist-3 multiparton distribution functions 
\begin{align}
F_{LT} (x_1,x_2),\ \ G_{LT} (x_1,x_2),\ \  
H_{LL}^\perp (x_1,x_2),\ \  H_{TT} (x_1,x_2),
\nonumber 
\end{align}
exist for tensor-polarized spin-1 hadrons.
These multiparton distribution functions are also interesting
for probing multiparton correlations in spin-1 hadrons.

\acknowledgments

The authors thank N. Sato for suggestions.
S. Kumano was partially supported by 
Japan Society for the Promotion of Science (JSPS) Grants-in-Aid 
for Scientific Research (KAKENHI) Grant Number 19K03830.
Qin-Tao Song was supported by the National Natural Science Foundation 
of China under Grant Number 12005191 and the Academic Improvement Project 
of Zhengzhou University.



\end{document}